\documentclass[prd,twocolumn,nofootinbib,notitlepage,aps,tightenlines,
preprintnumbers,amsmath,amssymb,amsfonts,showpacs,superscriptaddress]{revtex4-1}

\RequirePackage[colorlinks=true
,urlcolor=blue
,anchorcolor=blue
,citecolor=blue
,filecolor=blue
,linkcolor=blue
,menucolor=blue
,linktocpage=true
,pdfproducer=medialab
,pdfa=true
]{hyperref}

\usepackage{graphicx,epstopdf}
\usepackage{mathtools}
\usepackage{comment}
\usepackage{natbib,bm,bbm}
\usepackage[usenames,dvipsnames]{color}
\usepackage{slashed}    
\usepackage{xcolor}
\usepackage{multirow}
\usepackage{grffile}
\usepackage{physics}
\usepackage{dsfont}
\usepackage[]{hyperref}
\hypersetup{
	colorlinks,
	linkcolor={blue},
	linktocpage=true,
	citecolor={blue},
	urlcolor={blue}
}
\usepackage[capitalize]{cleveref}

\usepackage{float}
\usepackage{mathtools, nccmath}
\usepackage{lipsum}
\usepackage{cancel}

\def\be{\begin{equation}}
\def\ee{\end{equation}}
\def\ba{\begin{eqnarray}}
\def\ea{\end{eqnarray}}
\def\ge{\mathrel{\raise.3ex\hbox{$>$\kern-.75em\lower1ex\hbox{$\sim$}}}}
\def\la{\mathrel{\raise.3ex\hbox{$<$\kern-.75em\lower1ex\hbox{$\sim$}}}}

\def\simgt{\mathrel{\raise.3ex\hbox{$>$\kern-.75em\lower1ex\hbox{$\sim$}}}}
\def\simlt{\mathrel{\raise.3ex\hbox{$<$\kern-.75em\lower1ex\hbox{$\sim$}}}}

\newcommand{\nc}{\newcommand}

\nc{\gone}{\bar g_{\pi NN}^{(1)}}
\nc{\gzero}{\bar g_{\pi NN}^{(0)}}
\nc{\al}{\alpha}
\nc{\ga}{\gamma}
\nc{\de}{\delta}
\nc{\ep}{\epsilon}
\nc{\ze}{\zeta}
\nc{\et}{\eta}
\nc{\ka}{\kappa}
\nc{\rh}{\rho}
\nc{\si}{\sigma}
\nc{\ta}{\tau}
\nc{\up}{\upsilon}
\nc{\ph}{\phi}
\nc{\ch}{\chi}
\nc{\ps}{\psi}
\nc{\om}{\omega}
\nc{\Ga}{\Gamma}
\nc{\De}{\Delta}
\nc{\La}{\Lambda}
\nc{\Si}{\Sigma}
\nc{\Up}{\Upsilon}
\nc{\Ph}{\Phi}
\nc{\Ps}{\Psi}
\nc{\Om}{\Omega}
\nc{\ptl}{\partial}
\nc{\del}{\nabla}
\nc{\ov}{\overline}
\nc{\newcaption}[1]{\centerline{\parbox{15cm}{\caption{#1}}}}
\nc{\us}{U(1)$_S$}

\def\beq{\begin{equation}}
\def\eeq{\end{equation}}
\def\bmat{\begin{displaymath}}
\def\emat{\end{displaymath}}
\def\bear{\begin{eqnarray}}
\def\eear{\end{eqnarray}}
\def\ba{\begin{eqnarray}}
\def\ea{\end{eqnarray}}
\def\bery{\begin{array}}
\def\ery{\end{array}}
\def\bit{\begin{itemize}}
\def\eit{\end{itemize}}
\def\ben{\begin{enumerate}}
\def\een{\end{enumerate}}
\def\btab{\begin{tabular}}
\def\etab{\end{tabular}}
\def\btbl{\begin{table}}
\def\etbl{\end{table}}
\def\bfig{\begin{figure}[htb]}
\def\efig{\end{figure}}
\def\bpic{\begin{picture}}
\def\epic{\end{picture}}

\def\ga{\mathrel{\raise.3ex\hbox{$>$\kern-.75em\lower1ex\hbox{$\sim$}}}}
\def\la{\mathrel{\raise.3ex\hbox{$<$\kern-.75em\lower1ex\hbox{$\sim$}}}}
\def\gappeq{\mathrel{\rlap {\raise.5ex\hbox{$>$}}
{\lower.5ex\hbox{$\sim$}}}}
\def\lappeq{\mathrel{\rlap{\raise.5ex\hbox{$<$}}
{\lower.5ex\hbox{$\sim$}}}}

\def\gyr{{\rm \, G\kern-0.125em yr}}
\def\mev{{\rm \, Me\kern-0.125em V}}
\def\gev{{\rm \, Ge\kern-0.125em V}}
\def\tev{{\rm \, Te\kern-0.125em V}}

\def\slash#1{\rlap{\hbox{$\mskip 1 mu /$}}#1}%

\begin{document}


\preprint{UMN-TH-4319/24, FTPI-MINN-24-10}

\title{Chiral properties of the nucleon interpolating current and $\theta$-dependent observables}

\author{Yohei Ema}
\email{ema00001@umn.edu}
\affiliation{William I. Fine Theoretical Physics Institute, School of Physics and Astronomy, University of Minnesota,
Minneapolis, MN 55455, USA}
\affiliation{School of Physics and Astronomy, University of Minnesota,
Minneapolis, MN 55455, USA}

\author{Ting Gao}
\email{gao00212@umn.edu}
\affiliation{School of Physics and Astronomy, University of Minnesota,
Minneapolis, MN 55455, USA}

\author{Maxim Pospelov}
\email{pospelov@umn.edu}
\affiliation{William I. Fine Theoretical Physics Institute, School of Physics and Astronomy, University of Minnesota,
Minneapolis, MN 55455, USA}
\affiliation{School of Physics and Astronomy, University of Minnesota,
Minneapolis, MN 55455, USA}

\author{Adam Ritz}
 \email{aritz@uvic.ca}
\affiliation{Department of Physics and Astronomy, University of Victoria, Victoria BC V8P 5C2, Canada}

\date{\today}

\begin{abstract}
\noindent

We revisit the chiral properties of nucleon interpolating currents, and show that of the two leading order currents $j_1$ and $j_2$, only two linear combinations $j_1\pm j_2$ transform covariantly under the anomalous $U(1)_A$ symmetry. As a result, calculations of quantities which vanish by symmetry in the chiral limit may produce unphysical results if carried out with different linear combinations of the currents. This includes observables such as electric dipole moments, induced by the QCD parameter $\theta$, and the $\theta$-dependence of the nucleon mass. For completeness, we also exhibit the leading order results for nucleon electric dipole moments ($d_{n,p}$) induced by $\theta$, and the nucleon magnetic moments ($\mu_{n,p}$), when calculated using QCD sum rules for both the covariant choices of the nucleon interpolating current. The results in each channel, conveniently expressed as the ratios, $d_{n,p}/\mu_{n,p}$, are numerically consistent, and reflect the required physical dependence on $\theta$.

\end{abstract}
\maketitle

\section{Introduction}

Novel sources of $CP$-violation continue to be a primary target of searches for physics beyond the Standard Model, due to their potential role in clarifying the puzzle of the baryon asymmetry in the universe. Among many $CP$-odd sources, the $\theta$-term in QCD, ${\cal L} = \frac{\theta_G g_s^2}{32\pi^2}G^a_{\mu\nu} \widetilde{G}^{a\mu\nu}$, could contribute to a number of observables, but requires a fully nonperturbative analysis. Odd powers of $\theta_G$ break $P$ and $CP$ in flavor-diagonal channels. The experimentally verified absence of such observables way below the natural level implied by QCD has created a puzzle, known as the strong $CP$-problem~\cite{Graner:2016ses,Abel:2020pzs}. 

While the nonperturbative treatment of QCD in the hadronic regime is of course nontrivial, an important guide in elucidating the nature of $\theta$-dependent observables is the strong violation of chiral symmetry in the nonet of pseudoscalar Nambu-Goldstone mesons, $m_{\eta'} \gg m_{\pi,\eta_8}$. The nonvanishing of $m_{\eta'}$ in the chiral limit, $m_{u,d,s} \to 0$, was identified in \cite{Shifman:1979if} as the necessary and sufficient condition for $\theta$-dependence of physical observables. The dependence of the vacuum energy on $\theta$ (or equivalently the axion mass, if one promotes $\theta_G$ to a dynamical variable), the $\theta$-induced $CP$-odd pion-nucleon coupling, and the electric dipole moments (EDMs) of nucleons, all rely on a finite $m_\eta'$ in the chiral limit. 

 Among the nonperturbative approaches to $\theta$-dependent observables, lattice QCD promises to provide a systematic approach to the computation of the neutron EDM, $d_n(\theta)$, but results thus far are inconclusive and the program is ongoing; for a partial list of relevant papers, see {\em e.g.} Refs.~\cite{Aoki:1989rx,Guadagnoli:2002nm,Abramczyk:2017oxr,Dragos:2019oxn,Alexandrou:2020mds,Bhattacharya:2021lol,Bhattacharya:2022whc,Liang:2023jfj,He:2023gwp,Schierholz:2024var}. Approaches using chiral perturbation theory show relatively stable answers for the leading IR-singular terms~\cite{Crewther:1979pi,Dragos:2019oxn}. The QCD sum rule method, originating in Ref.~\cite{Shifman:1978bx}, is conceptually much closer to lattice QCD, and has also been employed to calculate $d_n(\theta)$ \cite{Pospelov:1999ha,Pospelov:1999mv,Pospelov:2005pr,Hisano:2012sc}, with results consistent with chiral estimates, but slightly smaller numerically. Nucleon magnetic moments have also been found within this method to be in reasonable agreement with observations \cite{Ioffe:1983ju}. 
 
 In this paper, our focus will be on a careful analysis of the chiral properties of the nucleon interpolating currents used in both lattice QCD and QCD sum rules, and to reassess the sum rules calculations of $d_n(\theta)$, as a concrete means of testing the symmetry-based constraints inferred from more general considerations. In this context, the sum rule approach, based on calculations of the operator product expansion (OPE) for hadronic current correlators in an external $\theta$ background, offers the advantage that many symmetries of the problem, such as chiral symmetry, and the chiral re-phasing invariance, can be made manifest at the quark-gluon level. These symmetries will allow us to determine physical choices for nucleon interpolating currents that ensure the required scaling of observables in the chiral limit. 

To be concrete, recall that the physical $\theta$ angle, $\bar\theta = \theta_G + \theta_m$, also includes the phase $\theta_m = {\rm Arg Det} {\cal M}_q$
of the quark mass matrix ${\cal M}_q$, and thus any physical dependence on $\theta$ necessarily vanishes in the chiral limit. This is conveniently observed within QCD itself by using the anomalous $U(1)_A$ symmetry to rotate away $\theta_G$, so that the physical phase is captured entirely by a complex singlet mass term. Restricting to the case of two light flavours, this term has the form 
\be
 {\cal L} = -m_*\bar{\theta}  (\bar{u}i\gamma_5 u + \bar{d} i\gamma_5d) + \frac{1}{2}m_*\bar{\theta}^2 (\bar{u}u + \bar{d} d) + \cdots,
 \label{basis}
\ee
where $m_* = m_u m_d/(m_u+m_d)$. It follows that any physical dependence on $\theta$ must vanish as $m_*\rightarrow 0$. This is immediately apparent in $CP$-even observables such as the topological susceptibility $d^2 E_{\rm vac}/d\bar{\theta}^2 = -m_* \langle 0| \bar{u}u + \bar{d} d| 0\rangle$, and the $\theta$-dependence of the nucleon mass $d^2 m_N/d\bar{\theta}^2 = -m_* \langle N| \bar{u}u + \bar{d} d| N\rangle$. 

$CP$-odd observables of considerable phenomenological interest first arise at linear order in $\theta$, such as nucleon EDMs and $CP$-odd pion nucleon couplings, and must also vanish in the $m_*\rightarrow 0$ limit. Our focus in this paper will be on the properties of nucleon interpolating currents that are required to ensure this behaviour. For example, we
 can write the most general interpolating current for neutrons that just involves the leading quark fields and no derivatives as follows,
\be
 j_n^\beta(x) = j_1(x) + \beta j_2(x), \label{interp}
\ee
where $\beta$ is a numerical coefficient, and the two currents with the quantum numbers of the neutron are given by $j_1(x){=} 2\ep_{ijk}(d_i^T \mathcal{C}\gamma_5 u_j)d_k$ and $j_2(x) {=} 2\ep_{ijk}(d_i^T \mathcal{C}u_j)\gamma_5d_k$ (see Sec.~\ref{sec:nucleon_chiral} for further details).
The notation in (\ref{interp}) reflects the fact that only $j_1$ is nonzero in the nonrelativistic limit, and thus the value of $\beta$ is apparently unimportant for generic observables in the neutron rest frame. However, the nonrelativistic limit for nucleons, encapsulated by the naive quark model, may not always be a good starting point for real life QCD, which corresponds to the limit of nearly massless quarks. This distinction proves to be important for $CP$-odd observables that are intrinsically sensitive to chirality-violating parameters such as $m_*$, and the choice of interpolating current deserves further scrutiny. Indeed, we will show below that only the choices $\beta=\pm 1$, namely
\be
 j_n^{\pm}(x) = j_1(x) \pm j_2(x),
\ee
are fully consistent when computing the leading dependence of observables on quantities, such as $\theta$, that transform under the anomalous $U(1)_A$ symmetry. Other choices allow for an unphysical dependence of observables in the chiral limit. For example, we show that away from these two special points, nucleon current correlators depend explicitly on $\theta$ in the $m_*\rightarrow 0$ limit, in contradiction with (\ref{basis}). 

Subtleties in the treatment of nucleon correlators in the chiral limit are well known, but are only important when studying chirally sensitive observables such as those dependent on $\theta$. It was highlighted in \cite{Pospelov:1999ha} that in the presence of $CP$-violation, the coupling of the physical nucleon state (represented by a spinor $v$) to the nucleon interpolator acquires an additional unphysical phase $\al$, where $\langle 0|j_n|N\rangle = \lambda e^{i\alpha_n \gamma_5/2} v$. This phase can mix magnetic and electric dipole structures, and complicates the extraction of physical observables from two-point correlation functions. As we discuss below, one can consider special tensor structures from which the phase $\alpha_n$ decouples, such as $\{ F\cdot \sigma\gamma_5, \slash{p}\}$ as considered in \cite{Pospelov:1999ha}, or explicitly calculate the phase as advocated for the specific approaches to computing $d_n(\theta)$ in lattice QCD \cite{Abramczyk:2017oxr}. The lack of chiral invariance for the generic nucleon interpolators also manifests in nontrivial mixing with $CP$-conjugate currents (denoted $i_1$ and $i_2$ in \cite{Pospelov:1999ha}), dependent on the unphysical combination $\theta_G - \theta_m$ orthogonal to $\bar\theta$. In this work, we will further argue that the chiral non-invariance of $j_n^\beta$ leads in fact to a generic and unphysical dependence on $\theta$ in the chiral limit unless $\beta=\pm1$.

We then proceed to systematically analyze the leading order results for the magnetic and electric dipole moments of nucleons using QCD sum rules for both consistent choices of the current interpolator $j_n^{\pm}$, extending earlier work \cite{Pospelov:1999ha,Hisano:2012sc}. We report new results for the $\beta =-1$ choice finding that $d_n(\theta)$ is consistent, both in sign and magnitude, with earlier estimates of $d_n(\theta)$ using $\beta =+1$ \cite{Pospelov:1999ha,Pospelov:1999mv,Pospelov:2005pr,Hisano:2012sc}. This analysis also allows us to directly relate $d_{n,p}$ to 
the nucleon magnetic dipole moment (MDM) $\mu_{n,p}$. From general principles, it is clear that one should expect the scaling $d_n \propto (\bar{\theta} m_*/m_n)\times \mu_n$, and determining a concrete coefficient in this relation is another goal of this paper. Since $\mu_{n,p}$ are reproduced rather reliably in the QCD SR approach \cite{Ioffe:1983ju}, and recently on the lattice \cite{Djukanovic:2023beb}, this may be considered as a useful/natural normalisation for the EDMs.

The remaining sections of this paper are organized as follows. In Sec.~\ref{sec:nucleon_chiral}, we define the nucleon interpolating currents, and illustrate their transformation under $U(1)_A$ in a general basis. We find that only the combinations $\beta = \pm1$ transform covariantly. 
 In Sec.~\ref{sec:chiral_limit}, we consider the chiral $m_*\rightarrow 0$ limit, and demonstrate the unphysical dependence of correlation functions on $\theta$ in the chiral limit unless $\beta=\pm1$. In Sec.~\ref{sec:EMDM}, we generalize earlier calculations of the nucleon EDMs using QCD sum rules for both covariant choices of the interpolating current $j^\pm_{n,p}$, and use alternate channels sensitive to the magnetic moment to express EDMs in the ratio $d_{n,p}/\mu_{n,p}$. We conclude by discussing the implications of our results for calculations of $\theta$-dependent observables, {\em e.g.} using the $j_1$ current, in Sec.~\ref{sec:discussion}.

\section{Nucleon currents and chirality}
\label{sec:nucleon_chiral}

In general, at lowest dimension, there are two independent nucleon interpolating currents (after applying the Fierz identity) that have the same
quantum numbers as the nucleons:
\begin{align}
	j_1^{a} &= 2\epsilon_{ijk}\left(d_i^{T} \mathcal{C} \gamma_5 u_j\right) q^{a}_k, \\
	j_2^{a} &= 2\epsilon_{ijk}\left(d_i^{T} \mathcal{C} u_j\right) \gamma_5 q^{a}_k,
\end{align}
where $i, j, k$ are the color indices and $\mathcal{C}$ is the charge conjugate matrix that satisfies
$(\gamma^\mu)^{T} \mathcal{C} = - \mathcal{C}\gamma^\mu$.
The index ``$a$" represents the isospin and $q^a = (u, d)^T$.
We note that $d_i^{T} \mathcal{C} \gamma_5 u_j = u_j^{T} \mathcal{C} \gamma_5 d_i$ and 
$d_i^{T} \mathcal{C} u_j = u_j^{T} \mathcal{C} d_i$,
and hence we can rewrite the currents as
\begin{align}
	j_1^{a} &= \epsilon_{ijk}\left(q_i^{bT}\epsilon_{bc}\mathcal{C}\gamma_5 q_j^{c}\right) q^{a}_k,\\
	j_2^{a} &= \epsilon_{ijk}\left(q_i^{bT}\epsilon_{bc}\mathcal{C} q_j^{c}\right) \gamma_5 q^{a}_k,
\end{align}
where $\epsilon_{ab}$ is the anti-symmetric tensor with $\epsilon^{12} = +1$ and hence $\epsilon_{12} = -1$.
This form makes it explicit that the diquark products inside the brackets 
are invariant under both chiral and vector $SU(2)$ transformations. 
This immediately leads to the conclusion that all linear combinations of the currents transform covariantly under the $SU(2)$ chiral and vector rotation.

Parametrizing a general linear combination of the two interpolation functions as
\begin{align}
	j_a^\beta = j_1^{a} + \beta j_2^{a},
\end{align}
we see that for the special choices of $\beta = \pm1$,
\begin{align}
	j_{a}^+ &= 2\epsilon_{ijk}\left[-\left(q_{iL}^{T} \mathcal{C}q_{jL}\right)q_{kL}^{a}
	+ \left(q_{iR}^{T}\mathcal{C}q_{jR}\right)q_{kR}^{a}\right],
	\label{eq:j+_chiral}
	\\
	j_a^{-} &= 2\epsilon_{ijk}\left[\left(q_{iR}^{T} \mathcal{C}q_{jR}\right)q_{kL}^{a}
	- \left(q_{iL}^{T}\mathcal{C}q_{jL}\right)q_{kR}^{a}\right],
\end{align}
where we suppress the isospin indices of the quark products,
and the subscripts $``L/R"$ indicate the projections onto left-/right-handed components.
This tells us that $j_a^\pm$ are covariant under the anomalous $U(1)_A$ transformation 
$q_i \to e^{i\theta_A \gamma_5}q_i$, while the current $j_a^\beta$ is not covariant 
under the $U(1)_A$ rotation for $\beta \neq \pm1$.
In general, the current transforms as
\begin{align}
	j_a^\beta
	\to \frac{1+\beta}{2}e^{3i\theta_A \gamma_5} j_a^+
	+ \frac{1-\beta}{2}e^{-i\theta_A \gamma_5}j_a^-.
\end{align}
Note, in particular, that the current $j_a^0$ which contains the unique non-relativistic structure, and so is widely used in lattice QCD computations, does not
transform covariantly under the $U(1)_A$ rotation.

We anticipate that this non-covariance for $\beta \neq \pm 1$ may complicate the extraction of physical quantities from nucleon correlators that depend sensitively on the realization of the anomalous chiral symmetry in QCD. 
To see this, recall that current correlators may be computed by
introducing an external fermionic source term $\eta_a$, with
\begin{align}
	\mathcal{L}_\eta = \mathcal{L} + \bar{\eta}_a j_a^\beta + (\mathrm{h.c.}),
\end{align}
where $\mathcal{L}$ is the original QCD Lagrangian (including $CP$-odd $\theta$ phases). The nucleon current correlator then follows from a second-order variation of action with respect to $\eta_a$.
If $j_a^\beta$ transforms covariantly under a $U(1)_A$ rotation, \emph{i.e.} if $\beta = \pm1$,
we can preserve the anomalous $U(1)_A$ symmetry by re-absorbing the chiral phases in the source $\eta_a$.
In other words, we can treat $\eta_a$ as a spurion to render the Lagrangian, including the source term, invariant.
On the other hand, if $j_a^\beta$ does not transform covariantly under the $U(1)_A$ rotation,
we cannot keep the whole Lagrangian, including the source term, invariant under the $U(1)_A$ rotation.\footnote{
	This naturally requires us to treat the external sources that couple to the $j_a^\pm$ components inside
	$j_a^\beta$ separately.
	In other words, we are required to introduce two distinct sources, so that
	$\mathcal{L}_\eta = \mathcal{L} + \bar{\eta}_a^\pm j_a^\pm + (\mathrm{h.c.})$, to maintain the invariance.
	It then follows that the correlators are defined by the chiral covariant currents $j_a^\pm$.
}
As a result, it is not guaranteed that the final correlators maintain the 
anomalous $U(1)_A$ symmetry of the original theory, for example being
independent of the unphysical phase combination $\theta_G - \theta_m$. Nor does it guarantee the restoration of $\bar\theta$-independence in the chiral $m_*\to 0$ limit.\footnote{
The approach introduced in \cite{Pospelov:1999ha,Pospelov:1999mv} to account for leading-order mixing with $CP$-conjugate currents $i_1 = \gamma_5 j_2$ and $i_2 = \gamma_5 j_1$ removes dependence on $\theta_G - \theta_m$, but may still induce an unphysical $\bar\theta$ dependence in the chiral limit unless $\beta = \pm1$.
}
This consideration naturally invites us to use the covariant currents $j_a^\pm$.

In the rest of this paper, we compute the nucleon correlators explicitly and confirm our general argument above;
the unphysical phase $\theta_G-\theta_m$ in general shows up in the correlators of $j_a^\beta$ with $\beta \neq \pm1$,
while only the physical combination $m_* \bar{\theta}$ appears in higher-point correlators of $j_a^\pm$ (after properly
subtracting the chiral phase of the two-point function; see Sec.~\ref{sec:EMDM}).

\section{Nucleon correlators in the chiral limit}
\label{sec:chiral_limit}

In Sec.~\ref{sec:nucleon_chiral}, we have seen that the lowest dimension nucleon interpolation currents are in general not covariant
under the $U(1)_A$ transformation, with the exception of two linear combinations, $j_a^\pm$ with $\beta = \pm1$.
In this section, we begin our investigation of the consequence of this non-covariance
by taking the chiral limit, $m_q \to 0$, with $m_*/m_q$ fixed.
In this limit, from the general properties of QCD, all dependence on $\theta$ should disappear from physical quantities,
as it can be rotated away by the $U(1)_A$ transformation of the quarks.
Despite this general expectation, as we see below, unphysical dependence on $\theta$ remains 
in the correlators of the currents for general choices of $\beta$.
The unphysical dependence disappears only for $\beta = \pm1$, indicating that only these choices of currents
produce physical results.

In the chiral limit, we take the QCD Lagrangian as
\begin{align}
	\mathcal{L} &= \bar{q}i\slashed{D} q
	- \frac{1}{4}G^{a}_{\mu\nu}G^{a\mu\nu}
	+ \frac{\theta_G \alpha_s}{8\pi}G_{\mu\nu}^{a}\tilde{G}^{a \mu\nu},
\end{align}
where $\tilde{G}^{a\mu\nu} = \epsilon^{\mu\nu\rho\sigma}G^{a}_{\rho\sigma}/2$ with $\epsilon^{0123} = +1$.
We define the electromagnetic part of the covariant derivative as $D_\mu = \partial_\mu + i e_q A_\mu$
with $e_u = 2e/3$ and $e_d = -e/3$. We have $e > 0$ with this convention.
We define the nucleon correlator as
\begin{align}
	\Pi_n^\beta (p) =  i\int d^4x\,e^{ip \cdot x}\langle 0 \vert \mathcal{T}\{j_n^\beta (x), \bar{j}_n^\beta(0)\}\vert 0 \rangle.
\end{align}
In the following, we compute correlator structures corresponding to the nucleon mass and EDM in the presence of $\theta$, employing the operator product expansion (OPE) with large $p^2 < 0$, as the first crucial step in constructing the QCD sum rule.

\subsection{Nucleon mass}

We begin our discussion with correlators that are often used for the calculation of the nucleon mass.
We first note that, as argued above, we can rotate away the gluonic $\theta$ term via a $U(1)_A$ transformation,
$q\to e^{i\theta_G m_*\gamma_5/2m_q}q$.\footnote{
    For brevity, in the rest of this subsection, we choose $m_u=m_d$ and thus $m_*/m_q = 1/2$ so that the chiral rotations of $u,d$ quarks are pure $U(1)_A$ transformations.
} This indicates that we can write down the (color-diagonal) quark propagator in the presence of $\theta_G$ as
\begin{align}
    S_q(\theta_G) = e^{i\theta_G \gamma_5/4} S_q(\theta_G = 0) e^{i\theta_G \gamma_5/4}.
\end{align}
The massless quark propagator is given at leading order by
\begin{align}
    S_q(\theta_G = 0) =\frac{i\slashed{x}}{2\pi^2 x^4} -\frac{\langle \bar{q}q\rangle}{12}.
\end{align}
Here $\langle \bar{q}q\rangle$ is short-hand notation for the vacuum condensate of quarks, $\langle 0|\bar{q}q|0\rangle$.
We then insert this expression into the nucleon correlator, simplify the Dirac matrix structures, and perform the Fourier transformation to momentum space. Correlators with an odd number of gamma-matrices, $\slashed{p}$ in this particular case, are explicitly $\theta$-independent at leading order.
However, chirality flipping Dirac structures, proportional to Dirac matrices $\mathbbm{1}$ or $\gamma_5$, acquire $\theta$-dependence. The leading order OPE terms are linear in the quark condensate and are given by: 
\begin{align}
    \label{eq:2point}
    \left.\Pi_n^\beta\right\vert_{\mathbbm{1},\gamma_5} &{=} 
    \frac{\langle \bar{q}q\rangle}{16\pi^2}p^2 \log\left(-\frac{p^2}{\mu^2}\right)
    \nonumber \\
    &\times
    (1{-}\beta)
    \left[6(1{+}\beta)e^{i\theta_G \gamma_5/2} + (1{-}\beta)e^{-\theta_G\gamma_5/2}\right].
\end{align}
This is a generalization of a well known result  \cite{Ioffe:1981kw,Leinweber:1995fn} for an arbitrary $\theta$ angle. A dual description of the same physics is achieved via a sum over nucleon states, including the excited states, $\propto 
\sum_i \lambda_i^2 e^{i\alpha_i\gamma_5}(\slashed{p}-m_i)^{-1}e^{i\alpha_i\gamma_5}$.
We see that, if the currents are covariant, $\beta = \pm1$, 
we have only one chiral phase.\footnote{
	The correlator vanishes at this order for $\beta = +1$.
	One can repeat the computation at ${\cal O}(\langle \bar{q}q\rangle^3)$
	and obtain the same conclusion that the correlator with $\beta = +1$ contains
	only one chiral phase.
}
We can then interpret the phase as the chiral phase of the nucleon states $\alpha_i$
that needs to be subtracted to obtain a physical result. In equivalent language, we can reabsorb this phase into the definition of the source $\eta_a$ so that the nucleon mass correlator does not depend on $\theta$.

On the other hand, if the currents are not covariant, $\beta \neq \pm1$,
the correlator contains two chiral phases, $e^{\pm i\theta_G \gamma_5/2}$, 
and we cannot absorb both phases in the overall chiral phase of the nucleon state.
We may choose the phase so that it absorbs the term linear in $\theta_G$,
but the term quadratic in $\theta_G$ remains. This would lead to the erroneous conclusion that the nucleon mass should acquire $\theta^2$-dependent contributions in the $m_*\to 0$ limit, which is entirely an artifact of using non-covariant currents.
For instance, for $\beta = 0$, the correlator can be expressed as
\begin{align}
	  \left.\Pi_n^0\right\vert_{\mathbbm{1},\gamma_5} &\propto
	  e^{5i\gamma_5\theta_G/28}\left(1-\frac{3}{49}\theta_G^2 + \cdots\right)e^{5i\gamma_5\theta_G/28},
\end{align}
where the dots indicate higher order terms in $\theta_G$.
We may absorb the phase $e^{5i\gamma_5\theta_G/28}$ into the nucleon state,
but the terms in the bracket, including the term of order $\theta_G^2$, will still contribute to the chirality flipping structure.
Therefore, if we use this expression to estimate
the nucleon mass, we obtain an unphysical dependence on $\theta$ even in the chiral limit.
This is inconsistent with the general constraint following from Eq.~\eqref{basis}, indicating that the calculation
based on the non-covariant currents, in general, is flawed. At a technical level, this occurs because the nucleon currents away from $\beta =\pm1$ contain ``built-in" flips of the quark chiralities $q_L\leftrightarrow q_R$ that persist in the chiral limit. 

\subsection{Nucleon EDM}

In the above subsection, we have seen that the nucleon mass term acquires unphysical dependence on $\theta$ 
in the chiral limit for general $\beta \neq \pm 1$. This raises concerns about the use of {\em e.g.} the ``non-relativistic'' $\beta=0$ current for calculation of any $\theta$-dependent nucleon observable. We should anticipate similar issues for $CP$-odd operators such as nucleon EDMs
that are intrinsically sensitive to the $\theta$-phases,
and we indeed confirm this expectation below.
In the following, we again focus on the chirality flipping part, as used in lattice QCD calculations of the neutron EDM~\cite{Aoki:1989rx,Guadagnoli:2002nm,Abramczyk:2017oxr,Dragos:2019oxn,Alexandrou:2020mds,Bhattacharya:2021lol,Bhattacharya:2022whc,Liang:2023jfj,He:2023gwp,Schierholz:2024var}.

In the chiral limit, with a background electromagnetic field, the quark propagator is given by
\begin{align}
	S_q &= \frac{i\slashed{x}}{2\pi^2 x^4}
	\nonumber \\
	&-\frac{\langle \bar{q}q\rangle}{12}\left(1+i\gamma_5\theta_G\frac{m_*}{m_q}\right)
	-\frac{\tilde{\chi}_q}{24}F\cdot \sigma
	\left(1 + i\gamma_5 \theta_G\frac{m_*}{m_q}\right),
\end{align}
where we have restored the dependence on $m_*/m_q$ for clarity, $F\cdot \sigma = F_{\mu\nu}\sigma^{\mu\nu}$,
and the vacuum condensate in the presence of the external electromagnetic field is parametrized as
$\langle \bar{q}\sigma_{\mu\nu}q\rangle_{F} = \chi_q F_{\mu\nu}\langle \bar{q}q\rangle
= \tilde{\chi}_q F_{\mu\nu}$. The quantity $\chi_q$ is the so-called magnetic susceptibility of the QCD vacuum introduced in \cite{Ioffe:1983ju}, while $\tilde{\chi}_q$ is introduced here for brevity. 
By focusing on the chirality flipping structures relevant to the sum rule (see the discussion around Eq.~\eqref{eq:double-pole} below) and retaining only the leading singular part, 
the correlator in the external field is given by
\begin{align}
	\left.\Pi_n^{\beta}\right\vert_{\{\slashed{p}, \{\slashed{p},F\cdot \sigma/2\}\}}
	&= -\frac{(1-\beta)^2 \tilde{\chi}_u}{96\pi^2}\log\left(-\frac{p^2}{\mu^2}\right),
	\\
	\left.\Pi_n^{\beta}\right\vert_{\{\slashed{p}, \{\slashed{p},iF\cdot \sigma \gamma_5/2\}\}}
	&= -\theta_G\frac{(1-\beta)^2 \tilde{\chi}_u}{96\pi^2}\frac{m_*}{m_u}\log\left(-\frac{p^2}{\mu^2}\right),
\end{align}
Notice that the dependence on the unphysical phase $\theta$ does not disappear in this expression.

In the context of lattice computations of the neutron EDM, Refs.~\cite{Abramczyk:2017oxr,Bhattacharya:2021lol} 
have proposed canceling the spurious phase by subtracting the corresponding phase computed via the two-point function (representing the chiral phase of the nucleon state), $\alpha_n^\beta$, from the phase of the three-point function. 
The chirality flipping part of the two-point function was computed in (\ref{eq:2point}), and upon linearization in $\theta$ takes the following form, 
\begin{align}
	\left.\Pi_n^{\beta}\right\vert_{\mathbbm{1},\gamma_5}
	&= \frac{\langle \bar{q}q\rangle}{16\pi^2}p^2 \log\left(-\frac{p^2}{\mu^2}\right) (1-\beta)
	\nonumber \\ 
	&\times \left[7+5\beta + i\gamma_5\theta_G\left(
	6(1+\beta)\frac{m_*}{m_d}-(1-\beta)\frac{m_*}{m_u}
	\right)\right],
\end{align}
where we retain only the leading-order terms with the logarithm.
From this expression, we can read off the chiral phase, acting on the nucleon mass operator, as
\begin{align}
	\alpha_n^\beta = \left[\frac{6(1+\beta)}{7+5\beta}\frac{m_*}{m_d} - \frac{1-\beta}{7+5\beta}\frac{m_*}{m_u}\right]\theta_G.
\end{align}
Following~\cite{Abramczyk:2017oxr,Bhattacharya:2021lol}, we may subtract this chiral phase from the three-point
function to obtain
\begin{align}
	&\left.\Pi_n^{\beta}\right\vert_{\{\slashed{p}, \{\slashed{p},iF\cdot \sigma\gamma_5/2\}\}}
	+ \alpha_n^\beta \times \left.\Pi_n^{\beta}\right\vert_{\{\slashed{p}, \{\slashed{p},F\cdot \sigma/2\}\}}
	\nonumber \\
	&~~~~~~~~= -\frac{\tilde{\chi}_u\theta_G}{16\pi^2}\log\left(-\frac{p^2}{\mu^2}\right)
	\times \frac{(1-\beta)^2(1+\beta)}{7+5\beta},
\end{align}
where we note the minus sign arising from commuting $\gamma_5$ with $\slashed{p}$,
resulting in $+\alpha_n^\beta$ instead of $-\alpha_n^\beta$ in the first line.

As one can observe, the removal of the unphysical phase does not occur in the EDM correlator for a generic choice $\beta$. Since physical quantities must be independent of $\theta$ in the chiral limit, it appears that the procedure outlined in~\cite{Abramczyk:2017oxr,Bhattacharya:2021lol} requires the use of $\beta=\pm1$ currents to ensure the cancelation of spurious $\theta$-dependence in nucleon EDMs.

\section{EDM and MDM sum rules for $\beta=\pm1$}
\label{sec:EMDM}

Thus far we have seen that calculations based on the non-covariant currents generically
induce spurious $\theta$ dependence even in the chiral limit.
As discussed in Sec.~\ref{sec:nucleon_chiral}, 
the interpolation functions with $\beta = \pm1$ are covariant under the $U(1)_A$ transformation.
This property allows us to define two distinct procedures to obtain correlators invariant under the $U(1)_A$ transformation
and thus free from unphysical $\theta$ dependence:
\begin{itemize}

\item Use the chirality conserving structure (with an odd number of $\gamma_\mu$) in the correlator.

\item Use the chirality flipping structure (with an even number of $\gamma_\mu$) in the correlator, and
subtract the chiral phase computed from the two-point correlator.

\end{itemize}
The former procedure was originally proposed in~\cite{Pospelov:1999ha}, as dependence on the chiral phase $\theta_A$ automatically cancels due to the gamma-matrix identity: 
\begin{equation}
    e^{i\alpha\gamma_5}({\rm odd\,number\,of\,\gamma_\mu})e^{i\alpha\gamma_5} = {\rm odd\,number\,of\,\gamma_\mu}.
\end{equation}
As a result, the EDM correlator structure proportional to $\{F\cdot\sigma \gamma_5,\slashed{p}\}$ is guaranteed to depend only on the physical combination $m_* \bar{\theta}$, and unphysical phases do not make an appearance for $\beta = \pm 1$. This method makes it possible to calculate EDM correlators without the need to consider the two-point functions and rotation angles $\alpha_n$. For a generic choice of $\beta$, Ref.~\cite{Pospelov:1999ha} suggested to add the specific admixture of $CP$-rotated currents $(i_1,i_2)$ that restore the invariance under the $U(1)_A$ rotation in this channel and guarantee $m_* \bar{\theta}$-proportionality of the OPE.

The second procedure (using the channel with an even number of $\gamma_\mu$) has been applied in lattice QCD computations 
of the neutron EDM~\cite{Abramczyk:2017oxr,Bhattacharya:2021lol}, with the $\beta = 0$ current choice. We would like to follow this path and calculate EDMs in the channel with an even number of $\gamma$ matrices, 
but with the important observation that we must
use the covariant currents $\beta = \pm 1$ to ensure physical dependence on $\bar\theta$.
Since the currents are covariant, the two- and three-point functions obtain the same chiral phase
after performing the chiral rotation of the quarks, and hence their difference is guaranteed to be independent
of the $U(1)_A$ rotation angle. This, again, leads to the dependence only on the physical combination $m_* \bar{\theta}$.

In the following, we confirm that the neutron EDM indeed depends only on the physical combination $m_* \bar{\theta}$
for both procedures, based on the QCD sum rule technique. Moreover, we observe that results obtained this way are consistent between the two different channels, using tensor structures with odd and even numbers of $\gamma$ matrices. 

We focus on the terms up to linear order in $m_q$ and the $\theta$-angles, and begin from the QCD Lagrangian
\begin{align}
	\mathcal{L} &= \bar{q}\left[i\slashed{D} - m_q\right] q
	\nonumber \\ &
	- \frac{1}{4}G^{a}_{\mu\nu}G^{a\mu\nu}
	- \theta_m m_* \bar{q}i\gamma_5 q + \frac{\theta_G \alpha_s}{8\pi}G_{\mu\nu}^{a}\tilde{G}^{a \mu\nu}.
\end{align}
Following the QCD sum rule approach, we compute the correlator of the nucleon interpolation current, given by
\begin{align}
	\Pi_n^{\pm}(p) = i\int d^4x\,e^{ip \cdot x}\langle 0 \vert \mathcal{T}\{j_n^\pm (x), \bar{j}_n^\pm(0)\}\vert 0 \rangle,
\end{align}
based on the OPE, 
and compare it with the phenomenological expression to extract the nucleon MDM and EDM.
To avoid sensitivity to IR divergences, it is convenient to use the first procedure (using chirality conserving structures) for $\beta = +1$,
and the second procedure (using chirality flipping structures) for $\beta = -1$, respectively. We discuss each of them in the following subsections.

\subsection{Sum rules for $\beta = +1$}

We begin with the QCD sum rules of the neutron MDM and EDM for $\beta = +1$
and focus on the chirality conserving part, as in~\cite{Pospelov:1999ha,Pospelov:1999mv,Hisano:2012sc}.
Since the current with $\beta = +1$ is covariant, the chiral phase automatically cancels when we focus on the chirality
conserving structure, leading to the dependence only on the physical combination $m_*\bar{\theta}$, as shown in the following.

In this case, Eq.~\eqref{eq:j+_chiral} tells us that only the chirality conserving part of the quark propagator
contributes to the correlator, and hence we can take
\begin{align}
	S_q = \frac{i\slashed{x}}{2\pi^2 x^4}
	- \frac{i e_q}{8\pi^2}\frac{x^\mu}{x^2}\tilde{F}_{\mu\nu}\gamma^\nu \gamma_5
	- \frac{i \tilde{\chi}_q}{24} m_* \bar{\theta}x^\mu F_{\mu\nu} \gamma^\nu\gamma_5,
\end{align}
where we keep only the leading order terms
contributing to the MDM and EDM.
Note that only the physical combination $m_*\bar{\theta}$ appears in this expression
since the chirality conserving part does not depend on the spurious chiral phase.
After some calculation, the relevant part of the correlator can be written as follows,
\begin{align}
	\Pi_n^+
	&= \frac{4e_d - e_u}{64\pi^4}p^2 \log\left(-\frac{p^2}{\mu^2}\right) \left\{\slashed{p},F\cdot \sigma\right\}
	\nonumber \\
	&- \frac{4\tilde{\chi}_d-\tilde{\chi}_u}{16\pi^2} m_* \bar{\theta}\log\left(-\frac{p^2}{\mu^2}\right)
	\left\{\slashed{p},iF\cdot \sigma \gamma_5\right\},
\end{align}
where we retain only the leading parts relevant for a Borel transformation. The transformed correlator is given by
\begin{align}
	\mathcal{B}\left[\Pi_n^+\right]_{\{\slashed{p},F\cdot \sigma\}}
	&= -\frac{4e_d - e_u}{64\pi^4}M^2,
	\\
	\mathcal{B}\left[\Pi_n^+\right]_{\{\slashed{p},iF\cdot \sigma \gamma_5\}}
	&= \frac{4e_d-e_u}{16\pi^2}m_* \bar{\theta}\chi \langle \bar{q}q\rangle,
\end{align}
where the subscripts denote the corresponding Dirac structures and we assume $\chi_q = e_q \chi$.
Remarkably, both the MDM and EDM depend on the linear combination, $4e_d - e_u$, 
that appears in the constituent quark model.\footnote{This property of the EDM correlator was noted in~\cite{Pospelov:1999ha}, while here we note that the same property holds for the MDM.}
On the phenomenological side of QCD sum rules, we represent the correlator, with a sum over hadron resonances and a continuum. For our leading order estimates below, we can neglect the continuum and single pole contributions, concentrating only on the leading nucleon double pole terms associated with the neutron ground state:
\begin{align}
	\Pi_n^+ &= -\frac{\vert \lambda_n\vert^2 m_n}{2(p^2 - m_n^2)^2}
	\left[\mu_n \left\{\slashed{p},F\cdot \sigma\right\}
	+ d_n \left\{\slashed{p},iF\cdot \sigma\gamma_5\right\}
	\right].
\end{align}
This expression includes both the MDM and EDM terms, and $\lambda_n$ denotes the coupling to the neutron state, $\langle 0|j_n^\pm|n\rangle = \lambda_n v$, up to an overall phase \cite{Pospelov:1999ha} which as noted above cancels in the chirality conserving channel. After the Borel transformation we obtain,
\begin{align}	
\label{MDMSR}
	\mathcal{B}\left[\Pi_n^+\right]_{\{\slashed{p},F\cdot \sigma\}}
	&= -\frac{\vert \lambda_n \vert^2 m_n}{2M^4}\mu_n e^{-m_n^2/M^2},
	\\
	\mathcal{B}\left[\Pi_n^+\right]_{\{\slashed{p},iF\cdot \sigma \gamma_5\}}
	&= -\frac{\vert \lambda_n \vert^2 m_n}{2M^4}d_n e^{-m_n^2/M^2}.
\end{align}
Taking the ratio to eliminate $\lambda_n$, we obtain
\begin{align}
	d_n = -\mu_n  m_*\bar{\theta} \frac{4\pi^2 \chi \langle \bar{q}q\rangle}{M^2}.
 \label{dn+}
\end{align}
As advertised, this result indeed depends only on the physical combination $m_*\bar{\theta}$, as required.

In writing the estimate in the form (\ref{dn+}), relating $d_n$ to $\mu_n$, one should also make sure that the estimate for the MDM is reasonably close to its measured value. Normalizing the MDM expression (\ref{MDMSR}) using the sum rule for $\slashed{p}$ to eliminate $\lambda_n$, one obtains the following expressions:
\begin{equation}
    \label{MDMresults}
    \mu_n = \frac{2}{m_n}\times\left(\frac43e_d - \frac13e_u\right);~~ \frac{\mu_n}{\mu_p} = -\frac23.
\end{equation}
The observed ratio of MDM is, famously, in agreement with the $-2/3$ value that follows from the constituent quark model and is also obtained using QCD sum rules at $\beta=+1$. 

The magnitudes of $\mu_{n,p}$ at leading order are within 50\% of the observed values. For example, for the neutron the prediction is $-8/3=-2.67$ in units of the nuclear Bohr magneton, while the observed value is $-1.91$. The estimate (\ref{MDMresults}) can be improved further upon the inclusion of the subleading OPE terms in both the MDM and $\slashed{p}$ channels. The subleading terms include gluon and quark condensate corrections. While the quark condensate corrections explicitly vanish for $\beta= +1$, the inclusion of the gluon corrections for the $\slashed{p}$ (see \cite{Leinweber:1995fn} and references therein) and MDM structures, calculated here, lead to the result:
\begin{equation}
    \frac{\mu_n}{e/(2m_n)} = -\frac83\times \frac{1+\frac{b}{24M^4}}{1+\frac{b}{4M^4}}\simeq  -2.05~{\rm at}~M=m_n.
\end{equation}
Here $b$ parametrizes the strength of the gluon vacuum condensate, $b\equiv (2\pi)^2\langle(\alpha_s/\pi) G_{\mu\nu}^aG_{\mu\nu}^a\rangle \sim 1.2\,{\rm GeV}^4 $. This result is indeed remarkably close to the observed value of the MDM, and the corrections do not spoil the $-2/3$ prediction for $\mu_n/\mu_p$. Therefore, we can be confident that the $\beta=+1$ sum rules perform at least as well as the $\beta =-1$ sum rules \cite{Ioffe:1983ju} in the MDM channel.

To obtain a numerical estimate for the EDM, we re-write the above result in terms of the pion mass, 
\begin{equation}
   F_\pi^2  m_\pi^2 = - (m_u+m_d)\langle \bar qq\rangle,
\end{equation}
with $F_\pi \simeq 93$\,MeV as this reduces the dependence on the normalization scale, 
\begin{align}
	d_n = \mu_n  \bar{\theta} m_\pi^2\times \frac{m_um_d}{(m_u+m_d)^2}\frac{4\pi^2 \chi F_\pi^2 }{M^2}.
 \label{EDMmpi}
\end{align}
Taking the Borel normalization scale to be $M=m_n$, with $m_u/m_d =0.48$, leads to the result 
\begin{equation}
    d_n|_{\beta=+1} \simeq 2\times 10^{-16}\,e\,{\rm cm}\times \bar{\theta}\times \left(\frac{|\chi|}{6\,{\rm GeV}^{-2}}\right).
\end{equation}
Although this is a leading order estimate, it is consistent with the result obtained in \cite{Pospelov:1999ha} which accounts for higher-order terms. Notably, its value is sensitive to the magnetic susceptibility $\chi$ of the QCD vacuum. Initial estimates \cite{Ioffe:1983ju,Belyaev:1984ic} put the value of $\chi$ close to $-6\, \rm GeV^{-2}$ ($-5.7\pm0.6\, \rm GeV^{-2}$ \cite{Belyaev:1984ic}), while later work based on considerations of the chiral anomaly in asymmetric kinematics and the pion pole dominance \cite{Vainshtein:2002nv}, estimates this quantity to be $\chi\sim - N_c/(4\pi^2F_\pi^2) \simeq -9\, \rm GeV^{-2} $. One should also note that available lattice studies \cite{Bali:2012jv} have found this quantity to be a factor of 2-to-3 smaller than Refs. \cite{Ioffe:1983ju,Belyaev:1984ic}, albeit with a higher normalization scale. Therefore, we conclude that the value of $\chi$ still provides the leading source of numerical uncertainty. Finally, for completeness, we also note that the proton EDM in this approach is given by $d_p(\bar\theta) = (-3/2)\times d_n(\bar\theta)$.

\subsection{Sum rules for $\beta = -1$}

We next consider the sum rules for the neutron MDM and EDM using $\beta = -1$ 
and focusing on the chirality flipping structure. Following Ioffe~\cite{Ioffe:1981kw}, this is the most widely used current in the QCD SR literature, including the MDM analysis of Ref.~\cite{Ioffe:1983ju}. However, the EDM has not previously been computed using this channel, or with this choice of current. In this approach, the unphysical chiral phase does not automatically cancel in the three-point function 
(that depends on $F_{\mu\nu}$), but can be subtracted by computing it directly from the two-point function
(that does not depend on $F_{\mu\nu}$).
Since $j_{a}^-$ is covariant under the $U(1)_A$ transformation,
this subtraction procedure \cite{Abramczyk:2017oxr,Bhattacharya:2021lol} defines a quantity that is invariant under the $U(1)_A$ transformation,
leading to dependence only on the physical combination $m_* \bar{\theta}$.

The relevant part of the quark propagator is given by
\begin{align}
	S_q &= \frac{i\slashed{x}}{2\pi^2 x^4}
	\nonumber \\
	&-\frac{m_q}{4\pi^2 x^2}\left(1-i\gamma_5 \theta_m \frac{m_*}{m_q}\right)
	-\frac{\langle \bar{q}q\rangle}{12}\left(1+i\gamma_5\theta_G\frac{m_*}{m_q}\right)
	\nonumber \\
	&
	- \frac{i e_q}{8\pi^2}\frac{x^\mu}{x^2}\tilde{F}_{\mu\nu}\gamma^\nu \gamma_5
	-\frac{\tilde{\chi}_q}{24}F \cdot \sigma
	\left(1 + i\gamma_5 \theta_G\frac{m_*}{m_q}\right)
	\nonumber \\
	&
	+\frac{e_q m_q}{32\pi^2}\log\left(-\mu_\mathrm{IR}^2 x^2\right)
	F\cdot \sigma \left(1-i\gamma_5 \theta_m \frac{m_*}{m_q}\right).
\end{align}
Notice that for the final term, the propagator perturbed by both $m_q$ and $F_{\mu\nu}$, is sufficiently infrared-singular to necessitate the introduction of the corresponding cutoff $\mu_{\rm IR}$.
As described above, we first compute the chiral phase of the two-point function, given by
\begin{align}
	\left.\Pi_n^{-}\right\vert_{\mathbbm{1},\gamma_5}
	&= \frac{\langle \bar{q}q\rangle}{4\pi^2}
	\left(1-i\gamma_5 \theta_G \frac{m_*}{m_u}\right)p^2 \log\left(-\frac{p^2}{\mu^2}\right)
	\nonumber \\
	&- \frac{m_u}{32\pi^4}\left(1+i\gamma_5 \theta_m \frac{m_*}{m_u}\right)p^4 \log\left(-\frac{p^2}{\mu^2}\right),
\end{align}
where the subscript denotes the Dirac structures we focus on.
By performing the Borel transformation, we obtain
\begin{align}
	\mathcal{B}\left[\Pi_n^{-}\right]_{\mathbbm{1},\gamma_5}
	&= -\frac{\langle \bar{q}q\rangle M^2}{4\pi^2}\left(1-i\gamma_5 \theta_G\frac{m_*}{m_u}\right)
	\nonumber \\
	&+ \frac{m_u M^4}{16\pi^4}\left(1+i\gamma_5 \theta_m \frac{m_*}{m_u}\right).
\end{align}
From this expression, we extract the chiral phase $\alpha_n^{-}$ as
\begin{align}
	\alpha_n^{-} = -\frac{m_*}{m_u}\theta_G - \frac{m_* M^2}{4\pi^2 \langle \bar{q}q\rangle}\bar{\theta}.
\end{align}
Note that the first term depends on both the physical and unphysical combinations of the phases,
$2\theta_G = \bar{\theta} + (\theta_G - \theta_m)$.

To compute the external field dependent three-point function, we note that
the correlator on the phenomenological side of the sum rule takes the form
\begin{align}
	\slashed{p}F\cdot \sigma \slashed{p} + m_n^2 F\cdot \sigma
	= \frac{1}{2}\left\{\slashed{p}, \{\slashed{p},F\cdot \sigma\}\right\}
	- (p^2 - m_n^2) F\cdot \sigma,
	\label{eq:double-pole}
\end{align}
for the MDM, and $F\cdot \sigma$ is replaced by $iF \cdot \sigma \gamma_5$ for the EDM.
Therefore, to focus on the double-pole contributions, 
we consider the Dirac structures $\{\slashed{p}, \{\slashed{p},F\cdot \sigma/2\}\}$ for the MDM
and $\{\slashed{p}, \{\slashed{p},iF\cdot \sigma \gamma_5/2\}\}$ for the EDM, respectively~\cite{Ioffe:1983ju}.
We denote the former structure as $``\mu"$ and the latter as $``\tilde{d}\,"$
for brevity, with the tilde indicating that the latter quantity is computed \emph{before} subtracting the chiral phase.
After some computation, these structures are given by
\begin{align}
	\Pi_n^{-}\vert_{\mu}
	&= -\frac{\tilde{\chi}_u}{24\pi^2}\log\left(-\frac{p^2}{\mu^2}\right) 
	\nonumber \\
	&+ \frac{m_u}{32\pi^4}\left[e_u  I(p^2) + e_d\log\left(-\frac{p^2}{\mu^2}\right)\right],
\end{align}
for the MDM, and
\begin{align}
	\Pi_n^{-}\vert_{\tilde{d}}
	&= -\frac{\tilde{\chi}_u}{24\pi^2}\frac{m_*\theta_G}{m_u} \log\left(-\frac{p^2}{\mu^2}\right)
	\nonumber \\
	&- \frac{m_*\theta_m}{32\pi^4}\left[e_u I(p^2) + e_d\log\left(-\frac{p^2}{\mu^2}\right) \right],
\end{align}
for the EDM (before the chiral phase subtraction), where
$I(p^2)$ is a function that encodes both UV and IR divergences, given explicitly as $I(p^2;\epsilon_\mathrm{IR},\epsilon_\mathrm{UV})$ in~\cite{Ema:2022pmo}. Here we only require its Borel transform, given by
\begin{align}
	\mathcal{B}[I(p^2)]
	= \log\left(\frac{M^2}{\mu_\mathrm{IR}^2}\right).
\end{align}
We then obtain
\begin{align}
	\mathcal{B}\left[\Pi_n^{-}\right]_{\mu}
	&= \frac{\tilde{\chi}_u}{24\pi^2} + \frac{e_u m_u}{32\pi^4}\left[\log\left(\frac{M^2}{\mu_\mathrm{IR}^2}\right)
	- \frac{e_d}{e_u}\right],
	\label{eq:MDM_beta-1}
	\\
	\mathcal{B}\left[\Pi_n^{-}\right]_{\tilde{d}}
	&=\frac{\tilde{\chi}_u}{24\pi^2}\frac{m_*}{m_u}\theta_G
	- \frac{e_u m_* \theta_m}{32\pi^4} \left[\log\left(\frac{M^2}{\mu_\mathrm{IR}^2}\right) - \frac{e_d}{e_u}\right].
\end{align}
Although the second term in Eq.~\eqref{eq:MDM_beta-1} is subdominant for the nucleon MDM,
it is important to obtain the physical combination, $m_* \bar{\theta}$, for the nucleon EDM
after subtracting the chiral phase.
Also, we distinguish $e_u \log\left({M^2}/{\mu_\mathrm{IR}^2}\right)$ and $e_d$
since they depend on different charges.

By subtracting the chiral phase $\alpha_n^{-}$ from the three-point functions, we obtain
\begin{align}
	&\mathcal{B}\left[\Pi_n^{-}\right]_{d}
	\equiv
	\mathcal{B}\left[\Pi_n^{-}\right]_{\tilde{d}}
	+ \alpha_n^{-} \times \mathcal{B}\left[\Pi_n^{-}\right]_{\mu}
	\nonumber \\
	&= -\left[\frac{\chi_u M^2}{96\pi^4} + \frac{e_u}{32\pi^4}
	\left( \log\left(\frac{M^2}{\mu_\mathrm{IR}^2}\right) - \frac{e_d}{e_u}\right)
	\right]m_* \bar{\theta},
\end{align}
where we again note the minus sign arising from commuting $\gamma_5$ with $\slashed{p}$,
resulting in $+\alpha_n^-$ instead of $-\alpha_n^-$ in the first line.
Notice that this now depends only on the physical combination, $m_*\bar{\theta}$, as expected.
On the phenomenological side of the sum rule, we have
\begin{align}
	\Pi_n^{-}
	&= -\frac{\vert \lambda_n \vert^2}{4(p^2 - m_n^2)^2}
	\left\{\slashed{p},\left\{\slashed{p},F\cdot \sigma\left(\mu_n + i\gamma_5 d_n\right)\right\}\right\},
\end{align}
for the MDM and EDM parts, 
where it is understood that this correlator holds \emph{after} rotating away the unphysical chiral phase.
Therefore we obtain our final result, after re-expressing $\vert \lambda_n \vert^2$ via the sum rule for the MDM,
\begin{align}
	d_n = -\mu_n  m_*\bar{\theta} \left[\frac{M^2}{4\pi^2 \langle \bar{q}q\rangle}
	{+} \frac{3}{4\pi^2 \chi \langle \bar{q}q\rangle} 
	\left(\log\left(\frac{M^2}{\mu_\mathrm{IR}^2}\right) {-} \frac{e_d}{e_u}\right)
	\right],
	\label{eq:EDM_beta-1}
\end{align}
where we denoted $\chi_u = e_u \chi$. 

We note the following qualitative features of this result:
\begin{itemize}
    \item As stated above, we see that the use of the covariant $\beta=-1$ current automatically leads to the correct dependence of EDM on $m_*\bar\theta$. 

    \item The $M^2/(4\pi^2 \langle \bar{q}q\rangle)$ term in (\ref{eq:EDM_beta-1}) results from the extraction of the $\alpha_n^-$ phase, and interferes destructively with the remaining terms. Using the leading order sum rule for the nucleon mass, $m_nM^2 = - 8\pi^2 \langle \bar{q}q\rangle$, known as the Ioffe formula approximation, this term can be rewritten as $-2/m_n$.

    \item Dependence on the infrared regulator $\mu_{\rm IR}$ means that the $\beta=-1$ result (\ref{eq:EDM_beta-1}) is less precise than for $\beta=+1$ due to the breakdown of the OPE. In particular, it is doubtful that using the scale separation one can calculate next-to-leading order corrections to (\ref{eq:EDM_beta-1}) without encountering power-like sensitivity to infrared scales. 

    \item The factor of $m_*$ in the numerator and $\langle \bar{q}q\rangle$ in the denominator form a combination that is far more sensitive to the normalization scale than the $\beta=+1$ result in Eq.~(\ref{EDMmpi}). 
    
\end{itemize}

With the caveats above, one can still make a parametric estimate of the EDM, by tentatively taking $M\sim 1$\,GeV, $\mu_{\rm IR} \sim 0.3$\,GeV, and $\langle \bar{q}q\rangle\simeq - (0.225\,{\rm GeV})^3$. Depending on the assumed value for $\chi$, that now enters in the denominator, numerical values for the EDM are in the range 
\begin{equation}
\label{EDMb=-1}
    d_n|_{\beta=-1} \sim (0.5\,\mbox{--}\,1.5)\times 10^{-16}\,e\,{\rm cm}\times \bar{\theta}.
\end{equation}
This result for the EDM indicates that we obtain the same sign for both $\beta=+1$ and $\beta=-1$ channels. This sign is also consistent with the chiral calculation, assuming the dominance of the chirally-enhanced $\log m_\pi$ contributions. The upper range of (\ref{EDMb=-1}) is for smaller values of $\chi \sim -3\,{\rm GeV}^{-2}$, at which point the $\beta =+1$ and $\beta = -1$ values for the EDM are approximately the same, and about two times smaller than chiral estimates for the $\log m_\pi$ contributions. 

\section{Discussion}
\label{sec:discussion}

The physical hadronic effects induced by the QCD vacuum angle $\bar\theta$ are subtle and depend sensitively on quantities that break chiral symmetry. Indeed, any matrix elements that depend on $\bar\theta$ also depend on $m_q$, rendering the quantitative impact at the per mille level when the quark mass is properly normalized on the hadronic mass, 
$m_*/m_n \sim O(10^{-3})$. This property follows directly from the QCD Lagrangian and the action of the anomalous $U(1)_A$ symmetry, but its implementation within modern methods that address hadronic/nucleon physics is far from straightforward.

Among a multitude of leading dimension nucleon interpolating currents $j_a^\beta$ parametrized by the angle $\beta$, only the choices $\beta=\pm1$ correspond to currents that transform covariantly under chiral rotations, {\em i.e.} preserving the same structure of the current, and acquiring an overall $e^{i\gamma_5\times {\rm phase}}$ phase. Importantly, one can then show that correlators of the corresponding currents $\Pi^{\pm}_n$ have the correct chiral properties and depend only on the physical combination $m_*\bar\theta$ for $m_* \ll m_n$, with $\theta_G+\theta_m=\bar\theta$. 

Conversely, we exhibited problematic features of correlators computed using other choices of currents, and in particular the $\beta=0$ choice often used in lattice QCD computations. The leading order OPE terms, that were calculated both for the two- and three-point functions, retain their $\theta$ dependence even in the chiral limit, $m_q\to 0$. This is because these currents, away from the $\beta =\pm 1$ point, contain spurious $q_L\leftrightarrow q_R$ chirality flips built into the interpolators that retain the phase dependence upon chiral rotations. Technically this manifests in the non-covariant transformation properties of such currents, and as a consequence $\Pi^{\beta \neq \pm}_n$ correlators retain unphysical phases dependence both in the mass and EDM/MDM channels even in the $m_q\to0$ limit. While these calculations are performed in the leading order of the OPE, it is nevertheless clear that this problem is a consequence of symmetries and not specific to this regime. As lattice QCD calculations approach the sensitivity required to see the physical effects of $\bar\theta$, use of the $\beta =\pm 1$ interpolating currents will ensure that the appropriate chiral extrapolation is under control. It is also worth emphasizing that the chiral covariance problem discussed here is unique to the $U(1)_A$ transformation. In contrast, $SU(2)$ chiral rotations, of the form $\exp(i\gamma_5 \tau^a\phi^a)$, will always result in a covariant transformation of all currents, due to the invariance of the diquark structure. Therefore, calculations of nucleon properties in the {\em e.g.} constant pion field background should produce physical results regardless of the choice of current. 

We also revisited EDM calculations for both the covariant $\beta = \pm 1$ choices of current, and calculated the EDM in parallel to the MDM. For $\beta= +1$ we reproduce the leading order result of Ref.~\cite{Pospelov:1999ha}, conveniently re-expressed as $d_{n,p}(\bar\theta)$ being proportional to the MDM, $\mu_{n,p}$. We note that this channel does reproduce the measured values of the MDM reasonably well, including the $\mu_n/\mu_p  =-2/3$ relation, and therefore $\mu_n$ can be used for normalization. 

A new calculation was presented using the $\beta=-1$ currents for the neutron and proton EDM. We utilized the channel with an even number of $\gamma$ matrices, and observed explicitly how the combination of the two- and three-point functions ({\em i.e.} explicit removal of the overall chiral phase) leads to physical results. We obtained a different, but nevertheless numerically consistent result for the neutron EDM. Extraction of quantitative predictions, and their systematic improvement within the QCD sum rules approach is problematic in this channel, as the leading order result already depends on the IR cutoff $\mu_{\rm IR}$. While this is a problem for the sum rules approach to nucleon correlators, it can be resolved within lattice QCD. Thus we hope that the procedure described here, using covariant $\beta=\pm1$ currents can be followed in lattice QCD computations of the nucleon EDMs.

We conclude by noting that for phenomenological purposes, it may be useful to revisit previous calculations of the neutron EDM due to higher-dimensional $CP$-odd sources such as the EDMs and chromo EDMs of quarks, using the approach pursued here of adding numerical stability by normalizing them on the MDM. Such sources are the primary targets in analyzing nucleon and atomic and molecular EDM sensitivity to new sources of $CP$ violation in nature \cite{Pospelov:2005pr}. In this context, we recall that while the inferred value of $\bar\theta$ is small possibly hinting at dynamical relaxation via the axion mechanism, the numerical value of $d_n(\bar\theta)$ still plays an important role in this context as the axion vacuum expectation value, $\bar\theta = \theta_{\rm ind}$, can be shifted away from zero in the presence of higher-dimensional sources.

\begin{acknowledgments}

Y.E., T.G. and M.P. are supported in part by U.S. Department of Energy Grant No. desc0011842, and the work of A.R. is supported by NSERC, Canada. T.G. and M.P. thank ECT* for support at the Workshop “EDMs: complementary experiments and theory connections” during which this work has been developed. 

\end{acknowledgments}

\bibliography{Refs.bib}

\begin{thebibliography}{27}%
\makeatletter
\providecommand \@ifxundefined [1]{%
 \@ifx{#1\undefined}
}%
\providecommand \@ifnum [1]{%
 \ifnum #1\expandafter \@firstoftwo
 \else \expandafter \@secondoftwo
 \fi
}%
\providecommand \@ifx [1]{%
 \ifx #1\expandafter \@firstoftwo
 \else \expandafter \@secondoftwo
 \fi
}%
\providecommand \natexlab [1]{#1}%
\providecommand \enquote  [1]{``#1''}%
\providecommand \bibnamefont  [1]{#1}%
\providecommand \bibfnamefont [1]{#1}%
\providecommand \citenamefont [1]{#1}%
\providecommand \href@noop [0]{\@secondoftwo}%
\providecommand \href [0]{\begingroup \@sanitize@url \@href}%
\providecommand \@href[1]{\@@startlink{#1}\@@href}%
\providecommand \@@href[1]{\endgroup#1\@@endlink}%
\providecommand \@sanitize@url [0]{\catcode `\\12\catcode `\$12\catcode
  `\&12\catcode `\#12\catcode `\^12\catcode `\_12\catcode `\%12\relax}%
\providecommand \@@startlink[1]{}%
\providecommand \@@endlink[0]{}%
\providecommand \url  [0]{\begingroup\@sanitize@url \@url }%
\providecommand \@url [1]{\endgroup\@href {#1}{\urlprefix }}%
\providecommand \urlprefix  [0]{URL }%
\providecommand \Eprint [0]{\href }%
\providecommand \doibase [0]{http://dx.doi.org/}%
\providecommand \selectlanguage [0]{\@gobble}%
\providecommand \bibinfo  [0]{\@secondoftwo}%
\providecommand \bibfield  [0]{\@secondoftwo}%
\providecommand \translation [1]{[#1]}%
\providecommand \BibitemOpen [0]{}%
\providecommand \bibitemStop [0]{}%
\providecommand \bibitemNoStop [0]{.\EOS\space}%
\providecommand \EOS [0]{\spacefactor3000\relax}%
\providecommand \BibitemShut  [1]{\csname bibitem#1\endcsname}%
\let\auto@bib@innerbib\@empty
\bibitem [{\citenamefont {Graner}\ \emph {et~al.}(2016)\citenamefont {Graner},
  \citenamefont {Chen}, \citenamefont {Lindahl},\ and\ \citenamefont
  {Heckel}}]{Graner:2016ses}%
  \BibitemOpen
  \bibfield  {author} {\bibinfo {author} {\bibfnamefont {B.}~\bibnamefont
  {Graner}}, \bibinfo {author} {\bibfnamefont {Y.}~\bibnamefont {Chen}},
  \bibinfo {author} {\bibfnamefont {E.~G.}\ \bibnamefont {Lindahl}}, \ and\
  \bibinfo {author} {\bibfnamefont {B.~R.}\ \bibnamefont {Heckel}},\ }\href
  {\doibase 10.1103/PhysRevLett.116.161601} {\bibfield  {journal} {\bibinfo
  {journal} {Phys. Rev. Lett.}\ }\textbf {\bibinfo {volume} {116}},\ \bibinfo
  {pages} {161601} (\bibinfo {year} {2016})},\ \bibinfo {note} {[Erratum:
  Phys.Rev.Lett. 119, 119901 (2017)]},\ \Eprint
  {http://arxiv.org/abs/1601.04339} {arXiv:1601.04339 [physics.atom-ph]}
  \BibitemShut {NoStop}%
\bibitem [{\citenamefont {Abel}\ \emph {et~al.}(2020)\citenamefont {Abel} \emph
  {et~al.}}]{Abel:2020pzs}%
  \BibitemOpen
  \bibfield  {author} {\bibinfo {author} {\bibfnamefont {C.}~\bibnamefont
  {Abel}} \emph {et~al.},\ }\href {\doibase 10.1103/PhysRevLett.124.081803}
  {\bibfield  {journal} {\bibinfo  {journal} {Phys. Rev. Lett.}\ }\textbf
  {\bibinfo {volume} {124}},\ \bibinfo {pages} {081803} (\bibinfo {year}
  {2020})},\ \Eprint {http://arxiv.org/abs/2001.11966} {arXiv:2001.11966
  [hep-ex]} \BibitemShut {NoStop}%
\bibitem [{\citenamefont {Shifman}\ \emph {et~al.}(1980)\citenamefont
  {Shifman}, \citenamefont {Vainshtein},\ and\ \citenamefont
  {Zakharov}}]{Shifman:1979if}%
  \BibitemOpen
  \bibfield  {author} {\bibinfo {author} {\bibfnamefont {M.~A.}\ \bibnamefont
  {Shifman}}, \bibinfo {author} {\bibfnamefont {A.~I.}\ \bibnamefont
  {Vainshtein}}, \ and\ \bibinfo {author} {\bibfnamefont {V.~I.}\ \bibnamefont
  {Zakharov}},\ }\href {\doibase 10.1016/0550-3213(80)90209-6} {\bibfield
  {journal} {\bibinfo  {journal} {Nucl. Phys. B}\ }\textbf {\bibinfo {volume}
  {166}},\ \bibinfo {pages} {493} (\bibinfo {year} {1980})}\BibitemShut
  {NoStop}%
\bibitem [{\citenamefont {Aoki}\ and\ \citenamefont
  {Gocksch}(1989)}]{Aoki:1989rx}%
  \BibitemOpen
  \bibfield  {author} {\bibinfo {author} {\bibfnamefont {S.}~\bibnamefont
  {Aoki}}\ and\ \bibinfo {author} {\bibfnamefont {A.}~\bibnamefont {Gocksch}},\
  }\href {\doibase 10.1103/PhysRevLett.63.1125} {\bibfield  {journal} {\bibinfo
   {journal} {Phys. Rev. Lett.}\ }\textbf {\bibinfo {volume} {63}},\ \bibinfo
  {pages} {1125} (\bibinfo {year} {1989})},\ \bibinfo {note} {[Erratum:
  Phys.Rev.Lett. 65, 1172 (1990)]}\BibitemShut {NoStop}%
\bibitem [{\citenamefont {Guadagnoli}\ \emph {et~al.}(2003)\citenamefont
  {Guadagnoli}, \citenamefont {Lubicz}, \citenamefont {Martinelli},\ and\
  \citenamefont {Simula}}]{Guadagnoli:2002nm}%
  \BibitemOpen
  \bibfield  {author} {\bibinfo {author} {\bibfnamefont {D.}~\bibnamefont
  {Guadagnoli}}, \bibinfo {author} {\bibfnamefont {V.}~\bibnamefont {Lubicz}},
  \bibinfo {author} {\bibfnamefont {G.}~\bibnamefont {Martinelli}}, \ and\
  \bibinfo {author} {\bibfnamefont {S.}~\bibnamefont {Simula}},\ }\href
  {\doibase 10.1088/1126-6708/2003/04/019} {\bibfield  {journal} {\bibinfo
  {journal} {JHEP}\ }\textbf {\bibinfo {volume} {04}},\ \bibinfo {pages} {019}
  (\bibinfo {year} {2003})},\ \Eprint {http://arxiv.org/abs/hep-lat/0210044}
  {arXiv:hep-lat/0210044} \BibitemShut {NoStop}%
\bibitem [{\citenamefont {Abramczyk}\ \emph {et~al.}(2017)\citenamefont
  {Abramczyk}, \citenamefont {Aoki}, \citenamefont {Blum}, \citenamefont
  {Izubuchi}, \citenamefont {Ohki},\ and\ \citenamefont
  {Syritsyn}}]{Abramczyk:2017oxr}%
  \BibitemOpen
  \bibfield  {author} {\bibinfo {author} {\bibfnamefont {M.}~\bibnamefont
  {Abramczyk}}, \bibinfo {author} {\bibfnamefont {S.}~\bibnamefont {Aoki}},
  \bibinfo {author} {\bibfnamefont {T.}~\bibnamefont {Blum}}, \bibinfo {author}
  {\bibfnamefont {T.}~\bibnamefont {Izubuchi}}, \bibinfo {author}
  {\bibfnamefont {H.}~\bibnamefont {Ohki}}, \ and\ \bibinfo {author}
  {\bibfnamefont {S.}~\bibnamefont {Syritsyn}},\ }\href {\doibase
  10.1103/PhysRevD.96.014501} {\bibfield  {journal} {\bibinfo  {journal} {Phys.
  Rev. D}\ }\textbf {\bibinfo {volume} {96}},\ \bibinfo {pages} {014501}
  (\bibinfo {year} {2017})},\ \Eprint {http://arxiv.org/abs/1701.07792}
  {arXiv:1701.07792 [hep-lat]} \BibitemShut {NoStop}%
\bibitem [{\citenamefont {Dragos}\ \emph {et~al.}(2021)\citenamefont {Dragos},
  \citenamefont {Luu}, \citenamefont {Shindler}, \citenamefont {de~Vries},\
  and\ \citenamefont {Yousif}}]{Dragos:2019oxn}%
  \BibitemOpen
  \bibfield  {author} {\bibinfo {author} {\bibfnamefont {J.}~\bibnamefont
  {Dragos}}, \bibinfo {author} {\bibfnamefont {T.}~\bibnamefont {Luu}},
  \bibinfo {author} {\bibfnamefont {A.}~\bibnamefont {Shindler}}, \bibinfo
  {author} {\bibfnamefont {J.}~\bibnamefont {de~Vries}}, \ and\ \bibinfo
  {author} {\bibfnamefont {A.}~\bibnamefont {Yousif}},\ }\href {\doibase
  10.1103/PhysRevC.103.015202} {\bibfield  {journal} {\bibinfo  {journal}
  {Phys. Rev. C}\ }\textbf {\bibinfo {volume} {103}},\ \bibinfo {pages}
  {015202} (\bibinfo {year} {2021})},\ \Eprint
  {http://arxiv.org/abs/1902.03254} {arXiv:1902.03254 [hep-lat]} \BibitemShut
  {NoStop}%
\bibitem [{\citenamefont {Alexandrou}\ \emph {et~al.}(2021)\citenamefont
  {Alexandrou}, \citenamefont {Athenodorou}, \citenamefont {Hadjiyiannakou},\
  and\ \citenamefont {Todaro}}]{Alexandrou:2020mds}%
  \BibitemOpen
  \bibfield  {author} {\bibinfo {author} {\bibfnamefont {C.}~\bibnamefont
  {Alexandrou}}, \bibinfo {author} {\bibfnamefont {A.}~\bibnamefont
  {Athenodorou}}, \bibinfo {author} {\bibfnamefont {K.}~\bibnamefont
  {Hadjiyiannakou}}, \ and\ \bibinfo {author} {\bibfnamefont {A.}~\bibnamefont
  {Todaro}},\ }\href {\doibase 10.1103/PhysRevD.103.054501} {\bibfield
  {journal} {\bibinfo  {journal} {Phys. Rev. D}\ }\textbf {\bibinfo {volume}
  {103}},\ \bibinfo {pages} {054501} (\bibinfo {year} {2021})},\ \Eprint
  {http://arxiv.org/abs/2011.01084} {arXiv:2011.01084 [hep-lat]} \BibitemShut
  {NoStop}%
\bibitem [{\citenamefont {Bhattacharya}\ \emph {et~al.}(2021)\citenamefont
  {Bhattacharya}, \citenamefont {Cirigliano}, \citenamefont {Gupta},
  \citenamefont {Mereghetti},\ and\ \citenamefont
  {Yoon}}]{Bhattacharya:2021lol}%
  \BibitemOpen
  \bibfield  {author} {\bibinfo {author} {\bibfnamefont {T.}~\bibnamefont
  {Bhattacharya}}, \bibinfo {author} {\bibfnamefont {V.}~\bibnamefont
  {Cirigliano}}, \bibinfo {author} {\bibfnamefont {R.}~\bibnamefont {Gupta}},
  \bibinfo {author} {\bibfnamefont {E.}~\bibnamefont {Mereghetti}}, \ and\
  \bibinfo {author} {\bibfnamefont {B.}~\bibnamefont {Yoon}},\ }\href {\doibase
  10.1103/PhysRevD.103.114507} {\bibfield  {journal} {\bibinfo  {journal}
  {Phys. Rev. D}\ }\textbf {\bibinfo {volume} {103}},\ \bibinfo {pages}
  {114507} (\bibinfo {year} {2021})},\ \Eprint
  {http://arxiv.org/abs/2101.07230} {arXiv:2101.07230 [hep-lat]} \BibitemShut
  {NoStop}%
\bibitem [{\citenamefont {Bhattacharya}\ \emph {et~al.}(2022)\citenamefont
  {Bhattacharya}, \citenamefont {Cirigliano}, \citenamefont {Gupta},
  \citenamefont {Mereghetti},\ and\ \citenamefont
  {Yoon}}]{Bhattacharya:2022whc}%
  \BibitemOpen
  \bibfield  {author} {\bibinfo {author} {\bibfnamefont {T.}~\bibnamefont
  {Bhattacharya}}, \bibinfo {author} {\bibfnamefont {V.}~\bibnamefont
  {Cirigliano}}, \bibinfo {author} {\bibfnamefont {R.}~\bibnamefont {Gupta}},
  \bibinfo {author} {\bibfnamefont {E.}~\bibnamefont {Mereghetti}}, \ and\
  \bibinfo {author} {\bibfnamefont {B.}~\bibnamefont {Yoon}},\ }\href {\doibase
  10.22323/1.396.0567} {\bibfield  {journal} {\bibinfo  {journal} {PoS}\
  }\textbf {\bibinfo {volume} {LATTICE2021}},\ \bibinfo {pages} {567} (\bibinfo
  {year} {2022})},\ \Eprint {http://arxiv.org/abs/2203.03746} {arXiv:2203.03746
  [hep-lat]} \BibitemShut {NoStop}%
\bibitem [{\citenamefont {Liang}\ \emph {et~al.}(2023)\citenamefont {Liang},
  \citenamefont {Alexandru}, \citenamefont {Draper}, \citenamefont {Liu},
  \citenamefont {Wang}, \citenamefont {Wang},\ and\ \citenamefont
  {Yang}}]{Liang:2023jfj}%
  \BibitemOpen
  \bibfield  {author} {\bibinfo {author} {\bibfnamefont {J.}~\bibnamefont
  {Liang}}, \bibinfo {author} {\bibfnamefont {A.}~\bibnamefont {Alexandru}},
  \bibinfo {author} {\bibfnamefont {T.}~\bibnamefont {Draper}}, \bibinfo
  {author} {\bibfnamefont {K.-F.}\ \bibnamefont {Liu}}, \bibinfo {author}
  {\bibfnamefont {B.}~\bibnamefont {Wang}}, \bibinfo {author} {\bibfnamefont
  {G.}~\bibnamefont {Wang}}, \ and\ \bibinfo {author} {\bibfnamefont {Y.-B.}\
  \bibnamefont {Yang}} (\bibinfo {collaboration} {\ensuremath{\chi}QCD}),\
  }\href {\doibase 10.1103/PhysRevD.108.094512} {\bibfield  {journal} {\bibinfo
   {journal} {Phys. Rev. D}\ }\textbf {\bibinfo {volume} {108}},\ \bibinfo
  {pages} {094512} (\bibinfo {year} {2023})},\ \Eprint
  {http://arxiv.org/abs/2301.04331} {arXiv:2301.04331 [hep-lat]} \BibitemShut
  {NoStop}%
\bibitem [{\citenamefont {He}\ \emph {et~al.}(2023)\citenamefont {He},
  \citenamefont {Abramczyk}, \citenamefont {Blum}, \citenamefont {Izubuchi},
  \citenamefont {Ohki},\ and\ \citenamefont {Syritsyn}}]{He:2023gwp}%
  \BibitemOpen
  \bibfield  {author} {\bibinfo {author} {\bibfnamefont {F.}~\bibnamefont
  {He}}, \bibinfo {author} {\bibfnamefont {M.}~\bibnamefont {Abramczyk}},
  \bibinfo {author} {\bibfnamefont {T.}~\bibnamefont {Blum}}, \bibinfo {author}
  {\bibfnamefont {T.}~\bibnamefont {Izubuchi}}, \bibinfo {author}
  {\bibfnamefont {H.}~\bibnamefont {Ohki}}, \ and\ \bibinfo {author}
  {\bibfnamefont {S.}~\bibnamefont {Syritsyn}},\ }in\ \href@noop {} {\emph
  {\bibinfo {booktitle} {{40th International Symposium on Lattice Field
  Theory}}}}\ (\bibinfo {year} {2023})\ \Eprint
  {http://arxiv.org/abs/2311.06106} {arXiv:2311.06106 [hep-lat]} \BibitemShut
  {NoStop}%
\bibitem [{\citenamefont {Schierholz}(2024)}]{Schierholz:2024var}%
  \BibitemOpen
  \bibfield  {author} {\bibinfo {author} {\bibfnamefont {G.}~\bibnamefont
  {Schierholz}},\ }\href@noop {} {\  (\bibinfo {year} {2024})},\ \Eprint
  {http://arxiv.org/abs/2403.13508} {arXiv:2403.13508 [hep-ph]} \BibitemShut
  {NoStop}%
\bibitem [{\citenamefont {Crewther}\ \emph {et~al.}(1979)\citenamefont
  {Crewther}, \citenamefont {Di~Vecchia}, \citenamefont {Veneziano},\ and\
  \citenamefont {Witten}}]{Crewther:1979pi}%
  \BibitemOpen
  \bibfield  {author} {\bibinfo {author} {\bibfnamefont {R.~J.}\ \bibnamefont
  {Crewther}}, \bibinfo {author} {\bibfnamefont {P.}~\bibnamefont
  {Di~Vecchia}}, \bibinfo {author} {\bibfnamefont {G.}~\bibnamefont
  {Veneziano}}, \ and\ \bibinfo {author} {\bibfnamefont {E.}~\bibnamefont
  {Witten}},\ }\href {\doibase 10.1016/0370-2693(79)90128-X} {\bibfield
  {journal} {\bibinfo  {journal} {Phys. Lett. B}\ }\textbf {\bibinfo {volume}
  {88}},\ \bibinfo {pages} {123} (\bibinfo {year} {1979})},\ \bibinfo {note}
  {[Erratum: Phys.Lett.B 91, 487 (1980)]}\BibitemShut {NoStop}%
\bibitem [{\citenamefont {Shifman}\ \emph {et~al.}(1979)\citenamefont
  {Shifman}, \citenamefont {Vainshtein},\ and\ \citenamefont
  {Zakharov}}]{Shifman:1978bx}%
  \BibitemOpen
  \bibfield  {author} {\bibinfo {author} {\bibfnamefont {M.~A.}\ \bibnamefont
  {Shifman}}, \bibinfo {author} {\bibfnamefont {A.~I.}\ \bibnamefont
  {Vainshtein}}, \ and\ \bibinfo {author} {\bibfnamefont {V.~I.}\ \bibnamefont
  {Zakharov}},\ }\href {\doibase 10.1016/0550-3213(79)90022-1} {\bibfield
  {journal} {\bibinfo  {journal} {Nucl. Phys. B}\ }\textbf {\bibinfo {volume}
  {147}},\ \bibinfo {pages} {385} (\bibinfo {year} {1979})}\BibitemShut
  {NoStop}%
\bibitem [{\citenamefont {Pospelov}\ and\ \citenamefont
  {Ritz}(1999)}]{Pospelov:1999ha}%
  \BibitemOpen
  \bibfield  {author} {\bibinfo {author} {\bibfnamefont {M.}~\bibnamefont
  {Pospelov}}\ and\ \bibinfo {author} {\bibfnamefont {A.}~\bibnamefont
  {Ritz}},\ }\href {\doibase 10.1103/PhysRevLett.83.2526} {\bibfield  {journal}
  {\bibinfo  {journal} {Phys. Rev. Lett.}\ }\textbf {\bibinfo {volume} {83}},\
  \bibinfo {pages} {2526} (\bibinfo {year} {1999})},\ \Eprint
  {http://arxiv.org/abs/hep-ph/9904483} {arXiv:hep-ph/9904483} \BibitemShut
  {NoStop}%
\bibitem [{\citenamefont {Pospelov}\ and\ \citenamefont
  {Ritz}(2000)}]{Pospelov:1999mv}%
  \BibitemOpen
  \bibfield  {author} {\bibinfo {author} {\bibfnamefont {M.}~\bibnamefont
  {Pospelov}}\ and\ \bibinfo {author} {\bibfnamefont {A.}~\bibnamefont
  {Ritz}},\ }\href {\doibase 10.1016/S0550-3213(99)00817-2} {\bibfield
  {journal} {\bibinfo  {journal} {Nucl. Phys. B}\ }\textbf {\bibinfo {volume}
  {573}},\ \bibinfo {pages} {177} (\bibinfo {year} {2000})},\ \Eprint
  {http://arxiv.org/abs/hep-ph/9908508} {arXiv:hep-ph/9908508} \BibitemShut
  {NoStop}%
\bibitem [{\citenamefont {Pospelov}\ and\ \citenamefont
  {Ritz}(2005)}]{Pospelov:2005pr}%
  \BibitemOpen
  \bibfield  {author} {\bibinfo {author} {\bibfnamefont {M.}~\bibnamefont
  {Pospelov}}\ and\ \bibinfo {author} {\bibfnamefont {A.}~\bibnamefont
  {Ritz}},\ }\href {\doibase 10.1016/j.aop.2005.04.002} {\bibfield  {journal}
  {\bibinfo  {journal} {Annals Phys.}\ }\textbf {\bibinfo {volume} {318}},\
  \bibinfo {pages} {119} (\bibinfo {year} {2005})},\ \Eprint
  {http://arxiv.org/abs/hep-ph/0504231} {arXiv:hep-ph/0504231} \BibitemShut
  {NoStop}%
\bibitem [{\citenamefont {Hisano}\ \emph {et~al.}(2012)\citenamefont {Hisano},
  \citenamefont {Lee}, \citenamefont {Nagata},\ and\ \citenamefont
  {Shimizu}}]{Hisano:2012sc}%
  \BibitemOpen
  \bibfield  {author} {\bibinfo {author} {\bibfnamefont {J.}~\bibnamefont
  {Hisano}}, \bibinfo {author} {\bibfnamefont {J.~Y.}\ \bibnamefont {Lee}},
  \bibinfo {author} {\bibfnamefont {N.}~\bibnamefont {Nagata}}, \ and\ \bibinfo
  {author} {\bibfnamefont {Y.}~\bibnamefont {Shimizu}},\ }\href {\doibase
  10.1103/PhysRevD.85.114044} {\bibfield  {journal} {\bibinfo  {journal} {Phys.
  Rev. D}\ }\textbf {\bibinfo {volume} {85}},\ \bibinfo {pages} {114044}
  (\bibinfo {year} {2012})},\ \Eprint {http://arxiv.org/abs/1204.2653}
  {arXiv:1204.2653 [hep-ph]} \BibitemShut {NoStop}%
\bibitem [{\citenamefont {Ioffe}\ and\ \citenamefont
  {Smilga}(1984)}]{Ioffe:1983ju}%
  \BibitemOpen
  \bibfield  {author} {\bibinfo {author} {\bibfnamefont {B.~L.}\ \bibnamefont
  {Ioffe}}\ and\ \bibinfo {author} {\bibfnamefont {A.~V.}\ \bibnamefont
  {Smilga}},\ }\href {\doibase 10.1016/0550-3213(84)90364-X} {\bibfield
  {journal} {\bibinfo  {journal} {Nucl. Phys. B}\ }\textbf {\bibinfo {volume}
  {232}},\ \bibinfo {pages} {109} (\bibinfo {year} {1984})}\BibitemShut
  {NoStop}%
\bibitem [{\citenamefont {Djukanovic}\ \emph {et~al.}(2023)\citenamefont
  {Djukanovic}, \citenamefont {von Hippel}, \citenamefont {Meyer},
  \citenamefont {Ottnad}, \citenamefont {Salg},\ and\ \citenamefont
  {Wittig}}]{Djukanovic:2023beb}%
  \BibitemOpen
  \bibfield  {author} {\bibinfo {author} {\bibfnamefont {D.}~\bibnamefont
  {Djukanovic}}, \bibinfo {author} {\bibfnamefont {G.}~\bibnamefont {von
  Hippel}}, \bibinfo {author} {\bibfnamefont {H.~B.}\ \bibnamefont {Meyer}},
  \bibinfo {author} {\bibfnamefont {K.}~\bibnamefont {Ottnad}}, \bibinfo
  {author} {\bibfnamefont {M.}~\bibnamefont {Salg}}, \ and\ \bibinfo {author}
  {\bibfnamefont {H.}~\bibnamefont {Wittig}},\ }\href@noop {} {\  (\bibinfo
  {year} {2023})},\ \Eprint {http://arxiv.org/abs/2309.06590} {arXiv:2309.06590
  [hep-lat]} \BibitemShut {NoStop}%
\bibitem [{\citenamefont {Ioffe}(1981)}]{Ioffe:1981kw}%
  \BibitemOpen
  \bibfield  {author} {\bibinfo {author} {\bibfnamefont {B.~L.}\ \bibnamefont
  {Ioffe}},\ }\href {\doibase 10.1016/0550-3213(81)90259-5} {\bibfield
  {journal} {\bibinfo  {journal} {Nucl. Phys. B}\ }\textbf {\bibinfo {volume}
  {188}},\ \bibinfo {pages} {317} (\bibinfo {year} {1981})},\ \bibinfo {note}
  {[Erratum: Nucl.Phys.B 191, 591--592 (1981)]}\BibitemShut {NoStop}%
\bibitem [{\citenamefont {Leinweber}(1997)}]{Leinweber:1995fn}%
  \BibitemOpen
  \bibfield  {author} {\bibinfo {author} {\bibfnamefont {D.~B.}\ \bibnamefont
  {Leinweber}},\ }\href {\doibase 10.1006/aphy.1996.5641} {\bibfield  {journal}
  {\bibinfo  {journal} {Annals Phys.}\ }\textbf {\bibinfo {volume} {254}},\
  \bibinfo {pages} {328} (\bibinfo {year} {1997})},\ \Eprint
  {http://arxiv.org/abs/nucl-th/9510051} {arXiv:nucl-th/9510051} \BibitemShut
  {NoStop}%
\bibitem [{\citenamefont {Belyaev}\ and\ \citenamefont
  {Kogan}(1984)}]{Belyaev:1984ic}%
  \BibitemOpen
  \bibfield  {author} {\bibinfo {author} {\bibfnamefont {V.~M.}\ \bibnamefont
  {Belyaev}}\ and\ \bibinfo {author} {\bibfnamefont {Y.~I.}\ \bibnamefont
  {Kogan}},\ }\href@noop {} {\bibfield  {journal} {\bibinfo  {journal} {Yad.
  Fiz.}\ }\textbf {\bibinfo {volume} {40}},\ \bibinfo {pages} {1035} (\bibinfo
  {year} {1984})}\BibitemShut {NoStop}%
\bibitem [{\citenamefont {Vainshtein}(2003)}]{Vainshtein:2002nv}%
  \BibitemOpen
  \bibfield  {author} {\bibinfo {author} {\bibfnamefont {A.}~\bibnamefont
  {Vainshtein}},\ }\href {\doibase 10.1016/j.physletb.2003.07.038} {\bibfield
  {journal} {\bibinfo  {journal} {Phys. Lett. B}\ }\textbf {\bibinfo {volume}
  {569}},\ \bibinfo {pages} {187} (\bibinfo {year} {2003})},\ \Eprint
  {http://arxiv.org/abs/hep-ph/0212231} {arXiv:hep-ph/0212231} \BibitemShut
  {NoStop}%
\bibitem [{\citenamefont {Bali}\ \emph {et~al.}(2012)\citenamefont {Bali},
  \citenamefont {Bruckmann}, \citenamefont {Constantinou}, \citenamefont
  {Costa}, \citenamefont {Endrodi}, \citenamefont {Katz}, \citenamefont
  {Panagopoulos},\ and\ \citenamefont {Schafer}}]{Bali:2012jv}%
  \BibitemOpen
  \bibfield  {author} {\bibinfo {author} {\bibfnamefont {G.~S.}\ \bibnamefont
  {Bali}}, \bibinfo {author} {\bibfnamefont {F.}~\bibnamefont {Bruckmann}},
  \bibinfo {author} {\bibfnamefont {M.}~\bibnamefont {Constantinou}}, \bibinfo
  {author} {\bibfnamefont {M.}~\bibnamefont {Costa}}, \bibinfo {author}
  {\bibfnamefont {G.}~\bibnamefont {Endrodi}}, \bibinfo {author} {\bibfnamefont
  {S.~D.}\ \bibnamefont {Katz}}, \bibinfo {author} {\bibfnamefont
  {H.}~\bibnamefont {Panagopoulos}}, \ and\ \bibinfo {author} {\bibfnamefont
  {A.}~\bibnamefont {Schafer}},\ }\href {\doibase 10.1103/PhysRevD.86.094512}
  {\bibfield  {journal} {\bibinfo  {journal} {Phys. Rev. D}\ }\textbf {\bibinfo
  {volume} {86}},\ \bibinfo {pages} {094512} (\bibinfo {year} {2012})},\
  \Eprint {http://arxiv.org/abs/1209.6015} {arXiv:1209.6015 [hep-lat]}
  \BibitemShut {NoStop}%
\bibitem [{\citenamefont {Ema}\ \emph {et~al.}(2022)\citenamefont {Ema},
  \citenamefont {Gao},\ and\ \citenamefont {Pospelov}}]{Ema:2022pmo}%
  \BibitemOpen
  \bibfield  {author} {\bibinfo {author} {\bibfnamefont {Y.}~\bibnamefont
  {Ema}}, \bibinfo {author} {\bibfnamefont {T.}~\bibnamefont {Gao}}, \ and\
  \bibinfo {author} {\bibfnamefont {M.}~\bibnamefont {Pospelov}},\ }\href
  {\doibase 10.1007/JHEP07(2022)106} {\bibfield  {journal} {\bibinfo  {journal}
  {JHEP}\ }\textbf {\bibinfo {volume} {07}},\ \bibinfo {pages} {106} (\bibinfo
  {year} {2022})},\ \Eprint {http://arxiv.org/abs/2205.11532} {arXiv:2205.11532
  [hep-ph]} \BibitemShut {NoStop}%
\end{thebibliography}%

\end{document}